\documentclass[pre,aps,amsmath,longbibliography,amsfonts,amssymb,twocolumn]{revtex4-1}
\newcommand{\etaval}{0.0724}
\newcommand{\etavalwe}{0.0724(5)}

\newcommand{\dpvalwe}{0.270(1)}

\newcommand{\zvalwe}{1.7170(5)}

\newcommand{\dvalwe}{0.2394(3)}

\newcommand{\bvalwe}{0.604(2)}
\newcommand{\nuvalwe}{2.53(1)}
\newcommand{\nuperpvalwe}{1.469(8)}
\newcommand{\phivalwe}{1.39(4)}
\newcommand{\ophival}{0.72}
\newcommand{\ophivalwe}{0.72(2)}

\newcommand{\pco}{0.574~41(4)}
\newcommand{\pctw}{0.569~475(25)}
\newcommand{\pcth}{0.566~62(3)}
\newcommand{\pcfo}{0.564~875(25)}
\newcommand{\pcfi}{0.563~79(2)}

\usepackage{graphicx,CJK}
\begin{document}
\begin{CJK*}{UTF8}{mj}
\title{Crossover behaviors in branching annihilating attracting walk}
	\author{Su-Chan Park (박수찬)}
\affiliation{The Catholic University of Korea, Bucheon 14662, Republic of Korea}
\begin{abstract}
	We introduce branching annihilating attracting walk (BAAW) in one dimension. 
	The attracting walk is implemented by a biased hopping in such a way that
	a particle prefers hopping to a nearest neighbor located 
	on the side where the nearest particle is found within the range of attraction. 
	We study the BAAW with four offspring by extensive Monte Carlo simulation.
	At first, we find the critical exponents of the 
	BAAW with infinite range of attraction, which are different from those 
	of the directed Ising (DI) universality class. 
	Our results are  consistent with the recent observation 
	[B. Daga and P. Ray, Phys. Rev. E {\bf 99}, 032104 (2019)]. 
	Then, by studying crossover behaviors, we show that as far 
	as the range of attraction is finite the BAAW belongs to the DI class.
	We conclude that the origin of non-DI critical behavior of the BAAW with infinite
	range of attraction is the long-range nature of the attraction.
\end{abstract}
\date{\today}
\maketitle
\end{CJK*}
\section{Introduction}
Various one dimensional models with two (sets of) 
symmetric absorbing states, which are also revealed as modulo-2 conservation in dynamics
of domain walls between two different absorbing states, and 
with finite range of interaction have been found to belong to the directed Ising (DI) 
universality class~\cite{GKvdT1984,TT1992,KP1994,M1994,BB1996,H1997,KC2003}.
An infinite barrier between symmetric absorbing states~\cite{HKPP1998} was
identified as an important condition for a model to belong to the DI universality class.
The validity of this assertion can be checked by studying what will happen if the condition
is not met.  Indeed, there are (at least) three different ways of breaking the condition and 
each way mediates distinct crossover from the DI class to the directed percolation (DP) 
universality class~\cite{BB1996,OM2008,PP2008PRE}.
The short-range interaction is also believed to be an important premise
because the long-range L\'evy flight 
changes the critical behavior~\cite{Vernon2001}.

Recently, a variant of the branching annihilating random walk
(BAW)~\cite{TT1992} was introduced in Ref.~\cite{DR2019}.
In this model, particles hop in a  biased manner and 
the direction of bias changes from configuration to configuration in such a way that 
hopping toward the nearest particle is preferred.
Daga and Ray~\cite{DR2019} assumed that the bias is a local interaction because
the BAW with odd number of offspring is still found to belong to the DP 
class even in presence of the bias.
This assumption sounds plausible in that a long-range interaction should
change the critical behavior of models in the DP class~\cite{J1981,G1982,Janssen1999,HH1999}.

However, the BAW with even number of offspring, which is a typical 
model in the DI class,
does not belong to the DI class any more if the bias is present~\cite{DR2019}.
If the bias is indeed a local interaction, then this study poses a challenge to
understanding the DI universality class. Two symmetric absorbing 
states with infinite barrier would not be sufficient to determine the DI class.

The purpose of this paper is to investigate whether the bias introduced in Ref.~\cite{DR2019} to the
BAW with even number of offspring has nothing to do with the long-range interaction. 
To this end, we come up with a modified version of Ref.~\cite{DR2019}
by introducing the range of attraction. We will present how the range of interaction
affects the critical behavior.

The structure of the paper is as follows. In Sec.~\ref{Sec:model},
we introduce the branching annihilating attracting walk (BAAW)
with finite range of attraction. In this model, hopping is biased only 
if a nearest particle is located within the range of attraction. When the range of
attraction is infinite, the BAAW is identical to the model in Ref.~\cite{DR2019}.
In Sec.~\ref{Sec:result}, we present our simulation results in two subsections.
In Sec.~\ref{Sec:inf}, we find the critical exponents of the BAAW with infinite
range of attraction (to be abbreviated as BAAW$_\infty$) more accurately than Ref.~\cite{DR2019}
by taking corrections to scaling into account. 
In Sec.~\ref{Sec:fin}, we present simulation results for the BAAW with finite range of
attraction, focusing on crossover behaviors.
Section~\ref{Sec:sum} summarizes our work with discussion.

\section{\label{Sec:model}Model}
The BAAW is defined on a one dimensional lattice of size $L$.
We always assume periodic boundary conditions.
A configuration $C$ at time $t$ is specified by the occupation number $a_i$ at
each site $i$ ($0\le i \le L-1$). 
$a_i$ can be either 1 or 0 and multiple occupancy is not allowed.
Each particle can either hop to one of its nearest neighbor sites
or branch $m$ particles. When a particle hops, the direction
is biased toward the closest particle if the closest particle is located within distance $R$.
If the distance to the closest particle is larger than $R$, then the hopping is symmetric.
Whenever two particles happen to occupy a same site, 
these two particles undergo a pair-annihilation reaction and are removed
from the system immediately.

Now we define the model more explicitly.
For ease of presentation, we introduce an indicator $I_i(x;C)$ 
for a given configuration $C$ as 
\begin{align}
	I_i(x;C) \equiv \delta\left(a_{i+x},1\right )\prod_{k=1}^{|x|-1}\delta\left (a_{i+ k},0\right )\delta\left (a_{i- k},0\right ),
	\label{Eq:indicator}
\end{align} 
where $\delta(a,b)$ is the Kronecker delta symbol that is one (zero) if $a=b$ ($a\neq b$).
In words, $I_i(x;C)$ is one if in configuration $C$
the nearest occupied site from site $i$ is $i+x$ irrespective of
whether site $i-x$ is occupied or not; and zero otherwise.
We introduce another indicator functions 
$\chi_i^\pm$,
\begin{align}
	\label{Eq:chipm}
	\chi_i^+ &\equiv \sum_{x=1}^R I_i(x;C),\quad
	\chi_i^- \equiv \sum_{x=1}^R I_i(-x;C).
\end{align}
Note that $\chi_i^+$ ($\chi_i^-$) is 1 if in configuration $C$ the nearest occupied site from $i$ is
within distance $R$ and is on the right (left) hand side of $i$; and 0 otherwise.
We will refer to $R$ as the range of attraction.
Notice that both $\chi^\pm$ can be 1 simultaneously if there is a positive $x \le R$ such that
$I_i(x;C) = I_i(-x;C)=1$.

Now we are ready to explain the dynamic rules of the BAAW.
Each particle can hop to one of its nearest neighbors with rate $p$ (hopping event)
or branch $m$ offspring with rate $1-p$ (branching event).
If a particle at site $i$, say, is decided to move in the hopping event,
then it hops to $i \pm 1$ with probability
\begin{align}
	H^\pm \equiv \frac{1}{2} \pm \epsilon (\chi_i^+ - \chi_i^-), 
\end{align}
where $0 \le \epsilon \le \frac{1}{2}$. 
A particle prefers hopping toward the closest particle if the closest particle is found within $R$. 

In the branching event, all sites $j$ with $0<|j-i|\le \lfloor m/2\rfloor$
will be given an offspring, 
where $\lfloor x \rfloor$ is the floor function (greatest integer not larger than $x$).
If $m$ is odd, then one more particle will be placed at site either $i+(m+1)/2$ or $i-(m+1)/2$,
which is chosen at random with equal probability.
If two particles happen to occupy the same site by dynamics
(either hopping or branching),
then they are annihilated together (typically represented by $A+A\rightarrow 0$) in no time.  

Notice that 
if either $\epsilon = 0$ or $R=0$, then the model is the same as the branching annihilating random 
walk~\cite{TT1992},
which belongs to the DI class if $m$ is even.
Also note that if $\epsilon>0$ and $R=\infty$, then the model is identical to that in 
Ref.~\cite{DR2019}. 
We fix $\epsilon=0.1$ in the next section, but
our preliminary simulations showed that other nonzero value of $\epsilon$ does not
alter the conclusion.

We use two different initial conditions. The first initial condition is such that
all sites (for finite $L$) are occupied. 
The second one is such that
only two consecutive sites are occupied in an infinite lattice ($L=\infty$).
In this case, the site index runs over all integers and the initial configuration 
can be represented as $a_0=a_1=1$ and $a_i=0$ for $i\neq 0, 1$.
We will refer to a Monte Carlo simulation with the first (second)
initial condition as the steady-state (dynamic) simulation.

In the steady-state simulation, we are interested in the density of occupied sites at time $t$,
which is defined as
\begin{align}
	\rho(t) =  \frac{\langle M_t \rangle}{L},
\end{align}
where $M_t = \sum_i a_i$ is the number of occupied sites at time $t$ and 
$\langle \ldots \rangle$ represents the average over all ensemble.
In actual simulation, $L$ is large enough that a finite-size effect is negligible.

In the dynamic simulation, we measure the mean number of particles $N(t)$ 
averaged over \emph{all} ensemble, the (survival) probability $S(t)$ that there is at 
least one particle in
the system at time $t$, and the spreading $R^2(t)$ 
averaged over \emph{surviving} ensemble.  These quantities are formally defined as
\begin{align}
	&N(t) \equiv \left \langle M_t \right \rangle,\nonumber\\
	&S(t) \equiv \left \langle \left (1 - \delta_{M_t,0}  \right ) \right \rangle,\nonumber\\
	&R^2(t) \equiv \left \langle \left (k_M-k_m \right )^2 \right \rangle /S(t),\\
	&k_M \equiv \left ( 1 - \delta_{M_t,0} \right )\max\{i|a_i=1\},\nonumber \\ 
	&k_m \equiv \left ( 1 - \delta_{M_t,0} \right )\min\{i|a_i=1\},
	\nonumber
\end{align}
where $\delta_{a,b}$ is the Kronecker $\delta$ symbol.

In all simulations, we measure quantities in question at a regularly spaced time points
on a logarithmic scale. To be specific,
the $i$th measurement is performed at $T_i$ defined as
\begin{align}
	T_i = 
	\begin{cases} 
		i , & 1\le i\le 40\\
		\lfloor 40 \times 2^{(i-40)/15} \rfloor, & 41 \le i \le 55,\\
		2 T_{i-15}, & 56 \le i.
	\end{cases}
	\label{Eq:measure}
\end{align}
Notice that 50 measurement points roughly correspond to a decade on a logarithmic scale.

\section{\label{Sec:result}Simulation Results}
\subsection{\label{Sec:inf}Infinite $R$}
We begin with presenting simulation results of the BAAW$_\infty$ with $m=4$ 
(four offspring). 
The BAAW$_\infty$ was already studied in Ref.~\cite{DR2019}, but we will provide
more accurate critical exponents and critical points.

We first present the results of the dynamic simulation.
At the critical point, the quantities in question exhibit power-law behaviors such as
\begin{align}
	N(t) \sim t^\eta,\quad
	S(t) \sim t^{-\delta'},\quad
	R^2(t) \sim t^{2/z},
\end{align}
where $\eta$, $\delta'$, and $z$ are critical exponents.
The critical exponents as well as the critical point are estimated
by analyzing effective exponents defined as
\begin{align}
	\eta_\text{e}(t;b) \equiv \frac{\ln \left [ N(t)/N(t/b)\right ]}{\ln b},\\
	-\delta'_\text{e}(t;b) \equiv \frac{\ln\left [ S(t)/S(t/b)\right ]}{\ln b},\\
	z_\text{e}(t;b) \equiv \frac{2 \ln b}{\ln\left [ R^2(t)/R^2(t/b)\right ] },
\end{align}
where $b>1$ is a constant.
If the system is at the critical point, then the effective exponents
should approach the corresponding critical exponents as $t\rightarrow \infty$.

To find the critical exponents 
accurately, 
we have to take corrections to scaling into account. If the long time
behavior of, for example, $N(t)$ at the critical
point is
\begin{align}
	N(t) = A t^\eta \left [ 1 + B t^{-\chi_\eta} + o(t^{-\chi_\eta}) \right ],
\end{align}
where $o(x)$ stands for all terms that approach zero faster than $x$ for small $x$,
then the effective exponent behaves asymptotically as
\begin{align}
	\eta_\text{e}(t;b) 
	\approx \eta - B \frac{b^{\chi_\eta}-1}{\ln b} t^{-\chi_\eta}.
	\label{Eq:asym_eta}
\end{align}

At the critical piont, $\eta_\text{e}$
shows a linear behavior for large $t$ when it is drawn as a function of $t^{-\chi_\eta}$.
If the system is in the active (absorbing) phase, 
then $\eta_\text{e}(t)$ should eventually veer up (down) as $t$ increases.
Hence, analyzing $\eta_\text{e}$ as a function of $t^{-\chi_\eta}$
with correct value of $\chi_\eta$ is important to find the critical point.
The critical exponent $\eta$, in turn, can be found by a linear extrapolation
of $\eta_\text{e}$ at the critical point.
Other exponents can be found in a similar manner.

\begin{figure}
	\includegraphics[width=\linewidth]{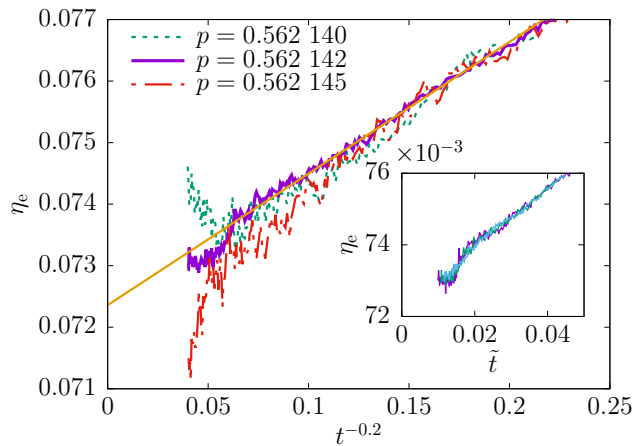}
	\caption{\label{Fig:eta} Plots of $\eta_\text{e}$ vs. $t^{-0.2}$
	for the BAAW$_\infty$ with $p=0.562~14$ (dotted green line), 
	$0.562~142$ (solid purple line), and $0.562~145$ (dot-dashed red line), top to bottom. 
	Here $b$ is set $16$.  The line shows the fitting of the data for
	$p=0.562~142$ with a linear function, which intersects the ordinate at $\etaval$. 
	Inset: Plots of $\eta_\text{e}$ vs. $\tilde t \equiv t^{-0.2}(b^{0.2}-1) /\ln b$ 
	for $p=0.562~142$ with $b=8$, 16, and 32.  Small $\tilde t$ (or large $t$) behavior 
	for different $b$ is hardly discernible, indicating that the corrections-to-scaling 
	exponent $\chi_\eta$ is estimated properly.
	}
\end{figure}
To find $\chi_\eta$, we analyze the corrections-to-scaling function
$Q_\eta$ defined as~\cite{Park2013,Park2014PRE}
\begin{align}
	Q_\eta(t;b,\chi) = \frac{\ln N(t) + \ln N\left (t/b^2\right ) - 2 \ln N(t/b)}{(b^\chi-1)^2},
\end{align}
whose asymptotic behavior at the critical point is
\begin{align}
	Q_\eta(t;b,\chi_\eta)\approx B t^{-\chi_\eta} .
	\label{Eq:Theta}
\end{align}
By analyzing how $Q_\eta$ behaves asymptotically, we can find
$\chi_\eta$ without prior knowledge of $\eta$. For consistency check of the estimated
$\chi_\eta$, we will depict $Q_\eta(t;b,\eta)$
for various $b$'s, which should behave identically in the asymptotic regime 
if $\chi=\chi_\eta$.
In a similar manner, we can also define $Q_{\delta'}$ and $Q_z$.
With our choice of measurement time points in Eq.~\eqref{Eq:measure}, 
it is convenient to set $b$ to be of the form  $2^n$ with positive integer $n$.

Now we present our results.
In actual simulations, the maximum
observation time is $T_{309} \approx 10^7$ and the number of independent runs for $p=0.562~142$, which 
will be shown to be the critical point,
is $2.5\times 10^8$.

\begin{figure}
	\includegraphics[width=\linewidth]{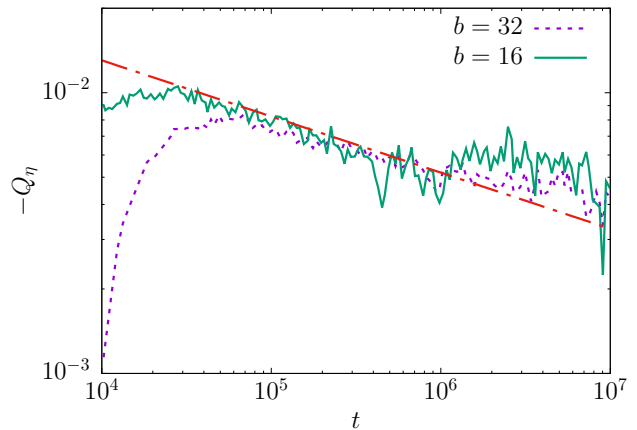}
	\caption{\label{Fig:qe} Double logarithmic plots of $-Q_\eta$ vs. $t^{-0.2}$
	for $b=16$ and 32 (left to right). 
	The dot-dashed line is $-B t^{-0.2}$ with $B$ ($B<0$) estimated in Fig.~\ref{Fig:eta}.
	}
\end{figure}
As a rule of thumb, $\eta$ is the best quantity to find the critical point in the dynamic
simulation. So we begin with the analysis of $\eta$.
In Fig.~\ref{Fig:eta}, we depict $\eta_\text{e}$ against 
$t^{-0.2}$ (that is, we use $\chi_\eta = 0.2$) for different $p$'s. Since $\eta_\text{e}$ for $p=0.562~14$ (0.562~145) eventually veers up (down)
for large $t$ and
$\eta_\text{e}$ for $p=0.562~142$ shows a linear behavior,
we conclude that the critical point is 
$p_0 = 0.562~142(3)$, where the number in parentheses indicates the
uncertainty of the last digit. 
Note that our estimate of $p_0$ (for $\epsilon=0.1$) is more accurate than 
Ref.~\cite{DR2019}.
Using a linear extrapolation, we find the critical exponent $\eta = \etavalwe$. For consistency
check, we depict 
$\eta_\text{e}$ vs. $t^{-0.2}(b^{0.2}-1)/\ln b$ at the critical point for $b=8$, 16, and 32 in the inset of Fig.~\ref{Fig:eta} . 
As expected from Eq.~\eqref{Eq:asym_eta}, these data lie on a single line.

We actually estimated $\chi_\eta$ from the analysis of $Q_\eta$ at $p=p_0$.
First note that $B$ in Eq.~\eqref{Eq:asym_eta} is negative as $\eta_\text{e}$ approaches
$\eta$ from above. Therefore, we draw $-Q_\eta(t)$ in Fig.~\ref{Fig:qe} 
for $b=16$ and 32 on a double logarithmic scale. 
For comparison, we also depict $-B t^{-0.2}$ with $B$ estimated 
when we found $\eta$ by a linear extrapolation. 
Consistently with Eq.~\eqref{Eq:Theta}, the asymptotic behavior 
for different $b$'s collapses into a single line, which concludes that $\chi_\eta$ is
indeed 0.2.

\begin{figure}
	\includegraphics[width=\linewidth]{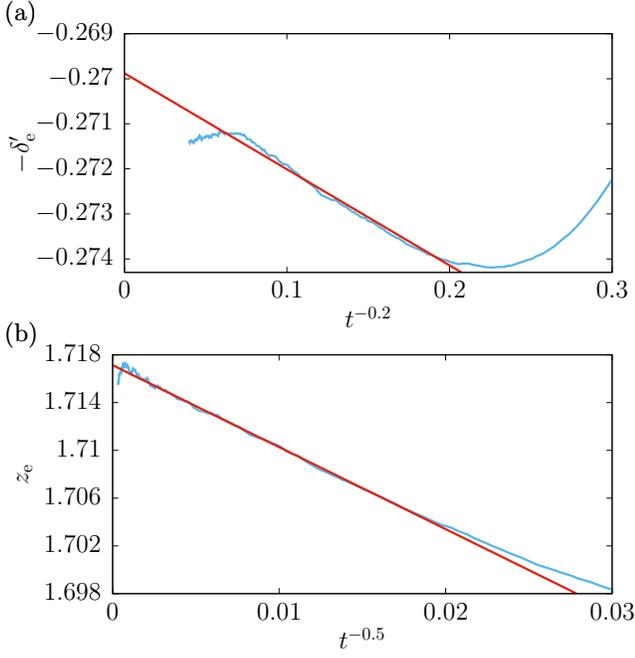}
	\caption{\label{Fig:dz} 
	Plots of the effective exponents at the critical point against $t^{-\chi}$.
	Here $b$ is set 16.
	The lines are results of fitting with a linear function.
	(a) The effective exponent $-\delta'_\text{e}$ 
	with $\chi=0.2$.
	(b) The effective exponent $z_\text{e}$ 
	with $\chi=0.5$.
	}
\end{figure}

In a similar fashion, we find $\delta'$ and $z$ using the corresponding effective exponent.
Figure~\ref{Fig:dz} shows the result of our analysis. We have first found 
that $Q_{\delta'} \sim t^{-0.2}$ and $Q_z \sim t^{-0.5}$ (details not shown here).
Linear extrapolations give $\delta' = \dpvalwe$ and $z = \zvalwe$.

Now we move on to the steady-state simulation. The density  $\rho(t)$
is expected to decay at the critical point $p_0$ as 
\begin{align}
	\rho(t) = A_\rho t^{-\delta} \left [ 1 + B_\rho t^{-\chi_\rho}
	+ o(t^{-\chi_\rho})\right ],
\end{align}
where $A_\rho$ and $B_\rho$ are constants.
As in the analysis of the dynamic simulation, we study the effective exponent
and the corrections-to-scaling function, defined as
\begin{align}
	-\delta_\text{e}
	&\equiv \frac{\ln \left [ \rho(t)/\rho(t/b)\right ]}{\ln b}
	\approx -\delta - B_\rho\frac{b^{\chi_\rho}-1}{\ln b}t^{-\chi_\rho},\\
	Q_\delta&\equiv \frac{\ln \rho(t)  + \ln \rho(t/b^2)- 2 \ln \rho(t/b)}{(b^{\chi_\rho}-1)^2}
	\approx B_\rho t^{-\chi_\rho},
	\nonumber
\end{align}
where we also presented the asymptotic behaviors at the critical point.

\begin{figure}
	\includegraphics[width=\linewidth]{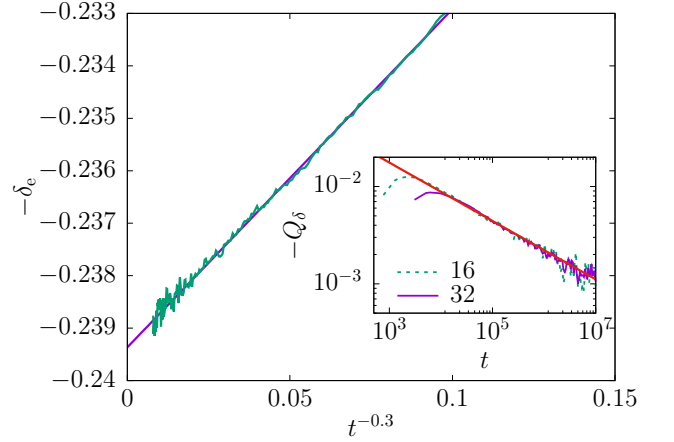}
	\caption{\label{Fig:del} 
	Plot of $-\delta_\text{e}$ vs. $t^{-0.3}$
	with $b=16$ at the critical point. 
	The line is the result of the linear fitting.
	Inset: Plot of $-Q_\delta$ vs. $t$ on a double logarithmic scale for $b=16$ and 32.
	The slope of the line is $-0.3$.
	}
\end{figure}

We simulated 1600 independent runs for the system 
of size $L = 2^{22}$ to $t=T_{309}\approx 10^7$ at the critical point $p_0$.
In Fig.~\ref{Fig:del}, we depict $-\delta_\text{e}$ against $t^{-0.3}$,
where we use $Q_\delta \sim t^{-0.3}$;
see the inset of Fig.~\ref{Fig:del}. 
By an extrapolation, we find $\delta =\dvalwe$. 
Notice that unlike the DI class $\delta$ is different from $\delta'$.

\begin{figure}[b]
	\includegraphics[width=\linewidth]{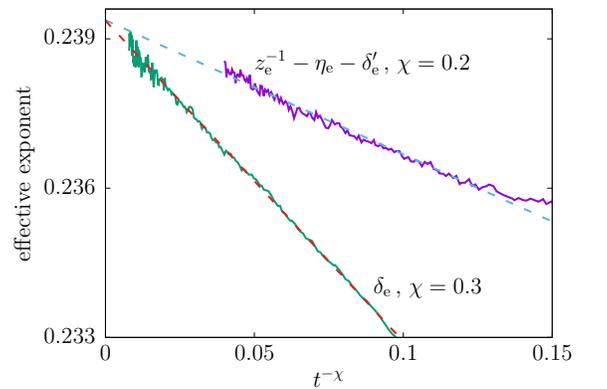}
	\caption{\label{Fig:hs} 
	Plots of effective exponents as a function of $t^{-\chi}$
	for $\delta_\text{e}$ with $\chi=0.3$ (bottom)
	and $z_\text{e}^{-1} -\eta_\text{e} - \delta'_\text{e}$ with $\chi=0.2$ (top).
	Here $b$ is 16.
	The dashed lines show extrapolations.
	}
\end{figure}

To check whether the scaling relation (see, e.g., Ref.~\cite{H2000})
\begin{align}
	\delta=  \frac{1}{z} -\eta  -\delta'
\end{align}
is satisfied, we depict the corresponding effective exponents in Fig.~\ref{Fig:hs},
which shows an excellent agreement.

Now we find $\beta$ by the steady-state simulation. To this end, we simulated
the system at $p = p_0 - \Delta$ 
with $\Delta=\Delta_n \equiv 2^{n/2} \times 10^{-4}$.
In actual simulations, $n$ ranges from $7$ to 18 ($n=7,8,\ldots,18$).
We will denote the steady-state density at $p = p_0 -\Delta_n$ by $\rho_n$.
To find the steady-state density, we simulated the system of size $L=2^{21}$ and 
averaged density for $t > 2 \Delta_n^{-2.53}$ (in fact,
2.53 is the estimated value of the critical exponent $\nu_\|$; see below).
The number of independent runs ranges from $500$ ($n=18$) to $4000$ ($n=7$).

To the leading order of corrections to scaling, we expect 
\begin{align}
	\rho_n = A_\beta \Delta_n^\beta \left [ 1 + B_\beta \Delta_n^\chi + o(\Delta_n^\chi)\right ],
\end{align}
where $A_\beta$ and $B_\beta$ are constants.
To estimate $\beta$, we also analyzed the effective exponent $\beta_\text{e}$
defined as
\begin{align}
	\beta_\text{e}(n;k) = \frac{\ln \left (\rho_{n+k}/\rho_n\right )}{\ln b}
	\approx \beta + B_\beta \frac{b^{\chi}-1}{\ln b} \Delta_n^\chi,
\end{align}
where $b = 2^{k/2}$ ($k=1,2,\ldots$).
To find $\chi$, we analyzed the corrections-to-scaling function~\cite{Park2020}
\begin{align}
	Q_\beta(n)&\equiv \frac{\ln \rho_{n+2k} + \ln \rho_n - 2 \ln \rho_{n+k}}{\left ( b^{\chi}-1 \right )^2}\approx B_\beta \Delta_n^\chi.
\end{align}

If the average is taken over $M$ configurations (recall that the measurement time is 
evenly distributed on a logarithmic scale), 
then we estimate the statistical error as
$e_n = 3\times \sigma/\sqrt{M}$, 
where $\sigma$ is the standard deviation of the measurement.
In our calculations, $M$ is generally larger than $10^4$.

Using the statistical error $e_n$ of the stationary-state density, 
we also estimate the error of $\beta_\text{e}$ by 
\begin{align}
	\beta^\pm = \frac{1}{\ln b} \ln \frac{\rho^0_{n+k} \pm e_{n+k}}{\rho^0_n \mp e_n},
	\quad \beta^- < \beta_\text{e} < \beta^+,
	\label{Eq:bee}
\end{align}
where $\rho_n^0$ is the average from simulations.

\begin{figure}
	\includegraphics[width=\linewidth]{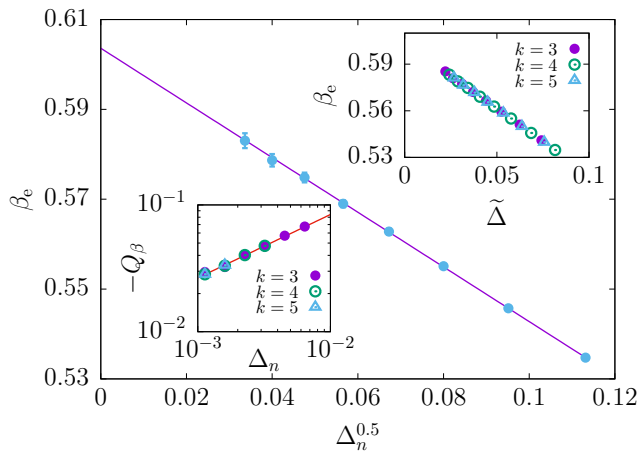}
	\caption{\label{Fig:beta} 
	Plots of  $\beta_\text{e}$ vs.  $\Delta_n^{0.5}$
	for $k=4$ ($b=2^{k/2}=4$). 
	The line is the result of a linear fitting.
	Inset: 
	(bottom) Plots of $-Q_\beta$ vs. $\Delta_n$ for $k=3,4,5$ on a double logarithmic scale.
	The slope of the line is 0.5;
	(top) plots of $\beta_\text{e}$ vs. $\widetilde \Delta$ 
	for $b=2 \sqrt{2}, 4$, and $4\sqrt{2}$, where
	$\widetilde \Delta \equiv \Delta_n^{0.5} (b^{0.5}-1)/\ln b$.
	}
\end{figure}
In Fig.~\ref{Fig:beta}, we show the behavior of $\beta_\text{e}$ obtained 
using $b=4$ (or $k=4$).
Data points of $\beta_\text{e}$ nicely lie on a line, when it is drawn as a 
function of $\Delta_n^{0.5}$. We actually
estimated $\chi$ from the behavior of
$Q_\beta(n)$, which shows $Q_\beta \sim \Delta_n^{0.5}$ irrespective of the value of $k$;
see the bottom inset of Fig.~\ref{Fig:beta}. In the top inset of Fig.~\ref{Fig:beta},
we also depict $\beta_\text{e}$ against $\widetilde \Delta \equiv \Delta_n^{0.5} (b^{0.5}-1)/\ln b$ for various $b$'s. As anticipated, $\beta_\text{e}$ as a function of $\widetilde \Delta$
does not exhibit any $b$ dependence for small $\widetilde \Delta$, which, in turn, supports that our choice of $\chi$ is legitimate.
Finally, by a linear extrapolation, we find $\beta =\bvalwe$.

In Table~\ref{Table:expo}, we summarized the critical exponents for the BAAW$_\infty$
with comparison to the DI critical exponents. 

\subsection{\label{Sec:fin}Crossover behavior for finite $R$}
Now we move on to the BAAW with finite range of attraction.
Before showing the simulation data, let us first ponder on the effect of
finite $R$.  When $R$ is finite and the system 
is very close to the critical point, the mean distance between particles can 
be much larger than $R$. Since the distance between a particle and its nearest 
particle is mostly larger than $R$, hopping becomes effectively unbiased.
Thus, the BAAW with finite $R$ should belong to the DI class.

Since the BAAW$_\infty$ was unequivocally shown not to belong to the DI class,
there should be a crossover behavior for sufficiently large but finite $R$.
This crossover is described by
a scaling function  (see, e. g., Refs.~\cite{LS1984,PP2006})
\begin{align}
	\rho(t;p,R) = t^{-\delta} \Psi\left [ t R^{-\mu_\|}, t^{1/\nu_\|} (p-p_0)\right],
	\label{Eq:cr_sc}
\end{align}
where $p_0$ is the critical point for the BAAW$_\infty$ we have found in the above; $\delta$ and $\nu_\|$ are the critical exponents
of the BAAW$_\infty$; and $\mu_\| = \nu_\| /\phi$ with $\phi$ to be the crossover exponent.

\begin{table}
	\caption{\label{Table:expo} Critical exponents of the BAAW$_\infty$ and
	the DI class. The exponents of the DI class are taken from Refs.~\cite{Park2013} and \cite{Park2020}.}
	\begin{ruledtabular}
		\begin{tabular}{lll}
			Exponent&BAAW$_\infty$ &DI\\
			\hline
			$\eta$&\etavalwe&0.0000(2)\\
			$\delta'$&\dpvalwe&0.2872(2)\\
			$z$&\zvalwe&1.7415(5)\\
			$\delta$&\dvalwe&0.2872(2)\\
			$\beta$&\bvalwe&1.020(5)\\
			$\nu_\perp$&\nuperpvalwe&2.04(1)\\
			$\nu_\|$&\nuvalwe&3.55(2)
		\end{tabular}
	\end{ruledtabular}
\end{table}
At the critical point $p_c(R)$ for finite $R$, the scaling function becomes
\begin{align}
	\rho(t;p_c,R) = t^{-\delta} \Psi\left [ t R^{-\mu_\|}, t^{1/\nu_\|} (p_c - p_0 )\right].
\end{align}
Since the scaling function at $p=p_c(R)$ for given $R$ should have a single scaling parameter,
$p_c(R)$ should behave as
\begin{align}
	p_c(R) - p_0 \sim R^{-1/\phi}.
\end{align}
Thus, the crossover exponent can be found by analyzing how $p_c$ behaves for large $R$.

To find the critical points, we exploit the fact that a plot of
$\rho(t) t^{\delta_\text{DI}}$, where $\delta_\text{DI}$ is the critical decay 
exponent of the DI class, should eventually veer up (down) 
if $p<p_c$ ($p>p_c$), where $\rho(t)$ is the density at time $t$ in
the steady-state simulation (details not shown here).
The critical points we have found are summarized in Table~\ref{Table:pc}.

\begin{figure}[b]
	\includegraphics[width=\linewidth]{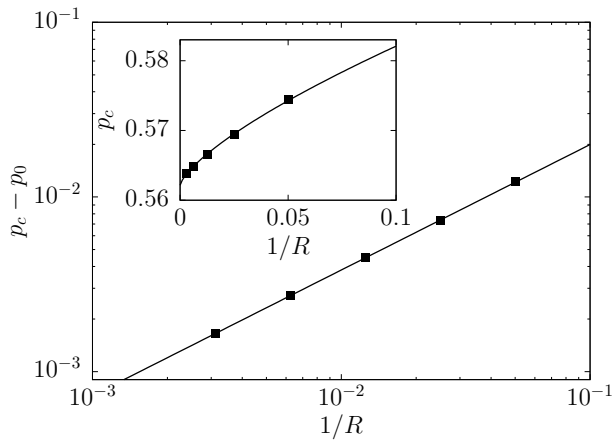}
	\caption{\label{Fig:phi} 
	Double logarithmic plot of $p_c-p_0$ vs.  $1/R$.
	A power-law fitting shows that $1/\phi \approx \ophival$,
	which is also depicted as a straight line.
	Inset: 
	Plot of $p_c$ vs. $1/R$. The curve shows the fitting function.
	}
\end{figure}

In Fig.~\ref{Fig:phi}, we show that
$p_c-p_0$ is well approximated by a power-law function with $1/\phi=\ophivalwe$, or $\phi=\phivalwe$, which together with $\nu_\|$ in Table~\ref{Table:expo} gives
$\mu_\| = 1.82(2)$.

\begin{figure}
	\includegraphics[width=\linewidth]{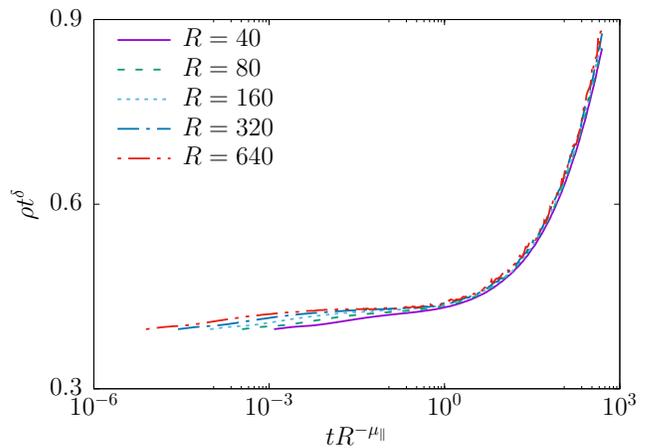}
	\caption{\label{Fig:cr} 
	Semilogarithmic plot of $\rho t^\delta$ vs.  $t R^{-\mu_\|}$ with $\mu_\|=1.82$
	for $R=40, 80, 160, 320$, and 640, right to left.
	Asymptotic behavior is well collapsed into a single curve.
	}
\end{figure}

If we set $p=p_0$ for finite $R$ in Eq.~\eqref{Eq:cr_sc}, then the scaling function becomes
\begin{align}
	\rho(t;p_0,R) t^{\delta} = \Psi\left (tR^{-\mu_\|},0\right ).
\end{align}
For consistency check, we present a scaling collapse in Fig.~\ref{Fig:cr},
using exponents obtained above.
The system size for Fig.~\ref{Fig:cr} is $L=2^{20}$ and
the number of independent runs ranges from 160 ($R=640$) to 800 ($R=40$).
As expected, all curves collapse onto a single curve,
which, in turn, supports that the BAAW with finite $R$ should belong to the DI class,
irrespective of how large $R$ is.

\begin{table}
	\caption{\label{Table:pc} Critical points of the BAAW for various values of the range
	of attraction $R$. Numbers in parentheses
	indicate errors in the last digits.}
	\begin{ruledtabular}
		\begin{tabular}{ll}
			$R$&$p_c$\\
			\hline
			%10&\pcz\\
			20&\pco\\
			40&\pctw\\
			80&\pcth\\
			160&\pcfo\\
			320&\pcfi\\
			%640&\pcs\\
			$\infty$&0.562~142(3)
		\end{tabular}
	\end{ruledtabular}
\end{table}
\section{\label{Sec:sum}Summary and Discussion}
To summarize, we studied the critical behavior of the BAAW
with four offspring.
We first estimated the critical exponents for the case $R=\infty$, which was
first reported in Ref.~\cite{DR2019}.
Our estimates of the critical point and the critical exponents are more accurate than 
Ref.~\cite{DR2019}.
Our finding reconfirms that the BAAW$_\infty$ does not belong to the directed Ising 
(DI) class.

To elucidate the origin of the different critical behavior, we studied the crossover,
occurring for finite $R$. We found that 
the BAAW with finite $R$ does belong to the DI universality class. We, furthermore,
estimated the crossover exponent $\phi =\phivalwe$, by studying the behavior of the critical 
point for large $R$. 
Since only the case of $R=\infty$ is different from the DI class,
we conclude that the different critical behavior of the BAAW$_\infty$
is due to the long-range attraction.

In contrast, the bias in the BAAW$_\infty$ was argued to be short-range interaction
in Ref.~\cite{DR2019}.
This argument is based on the fact that 
the BAAW with odd number of offspring is found to belong to the directed percolation  
universality class irrespective of whether $R=0$ or $R=\infty$, which 
is clearly different from the long-range jump dynamics~\cite{Janssen1999,HH1999,Vernon2001}.

However, the loss of long-range property of the bias for the case with odd $m$ in fact
originates from its dynamics.
To clarify this point, let us consider a situation, where there are only two particles in 
the (infinite) system and these particles are separated by $r$ which is assumed very large. 
For the bias (or, equivalently, the long-range attraction) to change the critical behavior, 
it is necessary that the long-time dynamics of this system should depend on $r$ as $r\rightarrow \infty$.
Otherwise, the bias is just a local interaction.

If the number of offspring is odd ($m=2\ell -1$), then each particle can undergo a 
(effectively) spontaneous annihilation by the combination of branching and pair-annihilation
dynamics ($A \rightarrow 2 \ell A \rightarrow 0$) before these two particles, more precisely two clusters
of particles, merge.
Thus, the survival probability that two clusters exist up to time $t$ does hardly depend on $r$ 
when $r \gg t$. At the critical point, this survival probability should decay as
$t^{-2\delta'}$ for $1 \ll t (\ll r)$.
In other words, the fate of each initial particle (more accurately the fate of each cluster) 
does not depend on the presence of other cluster and the long-time dynamics does not depend
on $r$ for large $r$.
Hence the bias becomes a (effectively) local interaction.

On the other hand, if the number of offspring is even as we have studied in this paper,
then the system cannot fall into the absorbing state
until initial two particles meet, even if the system is in the absorbing phase.
Hence, the characteristic time $\tau$ depends on $r$ and diverge as $r\rightarrow \infty$.
Since $\tau \sim r$ (biased diffusion) for infinite range of attraction and
$\tau \sim r^2$ (normal diffusion) for finite range of attraction as 
$r\rightarrow \infty$ (in the absorbing phase),
the long-time behavior differs only when the range of attraction is infinite.
This again explains why BAAW$_\infty$ is special.

Another direction of generalization of the BAAW$_\infty$ is
to associate the strength of the attraction with the distance to the nearest particle.
One possibility is to modify the probability that a particle hops toward the 
nearest particle by $\frac{1}{2} + \epsilon r^{-\sigma} $,
where $r$ is the distance to the nearest particle and $\sigma$ is a fixed number.
Clearly, the BAAW$_\infty$ corresponds to the case of $\sigma = 0$.
Our preliminary simulation shows that the critical exponents of the BAAW with even $m$ 
indeed does depend on $\sigma$,
as typical models with long-range interaction~\cite{Janssen1999,HH1999,Vernon2001}. 
This again shows that the different universal behavior
observed in Ref.~\cite{DR2019} is due to the long-range interaction. 
Detailed analysis will be published elsewhere~\cite{Pun}.

\begin{acknowledgments}
	This work was supported by the Basic Science Research Program through the
National Research Foundation of Korea~(NRF) funded by the Ministry of
Science and ICT~(Grant No. 2017R1D1A1B03034878) and by the Catholic University of Korea,
research fund 2019. 
The author furthermore thanks the Regional Computing Center of the
University of Cologne (RRZK) for providing computing time on the DFG-funded High
Performance Computing (HPC) system CHEOPS.
\end{acknowledgments}
\bibliography{Park}
\end{document}